\documentclass[showpacs,preprintnumbers,preprint,aps,amsmath,amssymb]{revtex4}
\usepackage{amsmath, amssymb, graphics,epsfig}
\newcommand{\mathsym}[1]{{}}

\def\lsim{\:\raisebox{-1.1ex}{$\stackrel{\textstyle<}{\sim}$}\:}
\def\gsim{\:\raisebox{-1.1ex}{$\stackrel{\textstyle>}{\sim}$}\:}
\def\10{$SO(10)$}
\def\21{SU(2) $\otimes$ U(1) }

\def\422{$SU(4) \otimes SU(2) \otimes SU(2)$}
\def\321{SU(3) $\otimes$ SU(2) $\otimes$ U(1)}
\def\lsim{\raise0.3ex\hbox{$\;<$\kern-0.75em\raise-1.1ex\hbox{$\sim\;$}}}
\def\gsim{\raise0.3ex\hbox{$\;>$\kern-0.75em\raise-1.1ex\hbox{$\sim\;$}}}

\newcommand{\ba}{\begin{array}}
\newcommand{\ea}{\end{array}}
\newcommand{\be}{\begin{equation}}
\newcommand{\ee}{\end{equation}}
\newcommand{\beqa}{\begin{eqnarray}}
\newcommand{\eeqa}{\end{eqnarray}}
\relax
\def\321{$SU(3)\times SU(2)\times U(1)$}
\def\mt{$\mu$-$\tau$ }

\begin{document}
\vspace*{2cm}
\title{Fermion Masses and Mixings in a $\mu$-$\tau$ symmetric $SO(10)$} 
\bigskip
\author{ Anjan S. Joshipura\footnote{anjan@prl.res.in}, Bhavik P.  
Kodrani\footnote{bhavik@prl.res.in} and Ketan M. Patel\footnote{kmpatel@prl.res.in}} \affiliation{Physical Research
Laboratory, Navarangpura, Ahmedabad-380 009, India. \vskip 2.0truecm}

\begin{abstract} 
\vskip 1.0 truecm 
The $\mu$-$\tau$ symmetry imposed on the neutrino mass matrix in the flavour basis is known to be quite predictive. We integrate this very specific neutrino symmetry into a more general framework based on the supersymmetric $SO(10)$ grand unified theory.  As in several other models, the fermion mass spectrum is determined by Hermitian mass matrices resulting from the renormalizable Yukawa couplings of the $16$-plet of fermions with the Higgs fields transforming as $10, \overline{126},120$ representations of the $SO(10)$ group. The $\mu$-$\tau$ symmetry is spontaneously broken through the $120$-plet. Consequences of this scheme are considered for fermion masses using both type-I and type-II seesaw mechanism. This scenario is shown to lead to a generalized CP invariance of the mass matrices and vanishing CP violating phases if the Yukawa couplings are invariant under the $\mu$-$\tau$ symmetry. Small explicit breaking of the $\mu$-$\tau$ symmetry is then shown to provide a very good understanding of all the fermion masses and mixing. Detailed fits to the fermion spectrum are presented in several scenarios. One obtains a very good fit to all observables in the context of the type-I seesaw mechanism but type-II seesaw model also provides a good description except for the overall scale of the neutrino masses. Three major predictions on the leptonic mixing parameters in the type-I seesaw case are  (1) the atmospheric mixing angle $\theta_{23}^{l}$ close to maximal, (2) $\theta_{13}^{l}$ close to the present upper bound and (3)
negative but very small Dirac CP violating phase in the neutrino oscillations. 
\end{abstract} 
\pacs{12.10-g,12.15Ff,14.60Pq,11.30Hv}

\maketitle
 
\section{Introduction}
There exist variety of theoretical frameworks/specific models \cite{rev} which try to account for the large atmospheric mixing angle observed more than a decade ago.  One class of theories attribute the maximal  atmospheric mixing to the presence of some underlying flavour symmetry. This would be a preferred alternative  if the deviation of the atmospheric mixing angle from maximality is constrained to be very small.  The simplest of such flavour symmetries is the \mt symmetry \cite{mt,hs,23,23as} which exchanges the mu and tau fields. This symmetry comes with an additional prediction that one of the three leptonic mixing angles namely, $\theta_{13}^{l}$ \cite{fnt} must be zero.

 \mt symmetry is predictive and simple but it appears to have two shortcomings.  Successful predictions follow only if it is an effective symmetry of the neutrino mass matrix in specific basis corresponding to a diagonal charged lepton mass matrix. The underlying flavour symmetry in general may not pick up this basis. Secondly, \mt symmetry  has been  proposed with a view of explaining the mixing angles in the leptonic sector alone. It would be  more desirable to have a symmetry providing  overall understanding of complete fermionic mass spectrum.  This can be done  using the grand unified theory as the underlying framework.
Various alternatives within such theories to simultaneously obtain small  mixing in the quark sector and large mixing among leptons have already been proposed \cite{rev,btau,asal}. 

The renormalizable theories based on the $SO(10)$ group are quite powerful in constraining the fermionic mass structures. The standard fermions are assigned to the $16$ dimensional representation of the $SO(10)$ group and they can obtain masses through symmetric couplings with $10$ and $\overline{126}$ and antisymmetric couplings with the 120 dimensional representation of the Higgs fields. The minimal $SO(10)$ model containing $10,126, \overline{126}$ and $210$ representations has been extensively studied \cite{btau,early,detailed1,detailed2,detailed3,bertolini,vissani2,splitsusy}.  
In this model, the  largeness of the atmospheric mixing angle gets related to the $b-\tau$ Yukawa unification if neutrinos obtain their masses through the type II \cite{typeII} seesaw mechanism \cite{seesaw}. This interesting observation in \cite{btau} led to many detailed investigations \cite{detailed3,vissani2,bertolini,splitsusy} which revealed the inadequacy of this simple picture. The supersymmetric version of the minimal model with type-II seesaw mechanism is constrained by two conflicting requirements. The overall neutrino mass scale is correctly reproduced in the model if the seesaw scale is about two to three orders of magnitude below the GUT scale. But the spectrum of the model in this case  does not allow gauge coupling unification.  Moreover, the type-II contribution to neutrino masses  does not always dominate over  the type-I contribution in the minimal model as would be required for the mechanism in \cite{btau} to go through. The conflict with the proton decay appears  in the minimal model even if the neutrinos obtain their masses through the type-I seesaw \cite{detailed3,vissani2,bertolini}. These problems have led to studies of the non-minimal models containing an additional 120-plet of Higgs \cite{charan120,grimus1,grimus2}. Theoretical understanding of the largeness of the atmospheric mixing angle gets lost in all these approaches although one can choose the parameters to obtain the observed value.

It would be welcome to integrate \mt symmetry into the grand unified framework. This has been done for the $SU(5)$ model in \cite{su5moh}. We do so here in the more predictive $SO(10)$ framework.  There are several motivations for unifying  $SO(10)$ and \mt symmetry.  Rather than remaining a leptonic symmetry, such symmetry would provide constraining picture of both quark and lepton spectrum. Role of this symmetry in the description of the quark mixing is already discussed in \cite{23,23as}.  In addition, it can provide additional constraints and reduces the number of the Yukawa couplings which describe fermion masses. Some examples of models unifying $SO(10)$ with other discrete symmetries can be found in \cite{discrete}.

We investigate the consequences of imposing a generalized \mt symmetry exchanging the second and the third generation fields on a renormalizable $SO(10)$ model. We deviate from the minimal model and add a 120 plet. This plays a crucial role in generating CP violation and the \mt symmetry breaking.  Fermion masses with 120 plet have been discussed  in several earlier works \cite{charan120, grimus1,grimus2,charan2,1202,1203,mohparity,yangwang}. Following \cite{mohparity,grimus1,grimus2,yangwang} we impose also the  parity symmetry which leads to Hermitian mass matrices for all fermions thereby reducing  the number of parameters compared  to more general models. All the fermion masses and mixing are described
in our approach in terms of  14 (15)  real parameters in case of type-II (type-I) seesaw mechanism.  They provide an excellent description of fermion masses and mixing in contrast to a general model employing $10+120+\overline{126}$ Higgs fields which needs \cite{mohparity}  31  parameters in the fermionic sector. Moreover, the (near) maximality of the atmospheric mixing and smallness of the angle $\theta_{13}^{l}$ get related here to the approximately broken \mt symmetry.

We define our model implementing \mt symmetry and discuss its consequences in the next section. Section(III) presents numerical fits both in case of the type-II and type-I seesaw mechanism and discusses various predictions. Last section contains a summary.
\section{\mt symmetric $SO(10)$}
 If  \mt symmetry is to be integrated with grand unification then a more general
symmetry which exchanges the second and  third generations of fermions should be imposed. 
Consequences of this generalization were  first considered in \cite{23}. It was subsequently noted \cite{23as} that this generalization automatically leads to understanding of why Cabibbo angle is larger than other two angles and a mild breaking of this symmetry was shown to lead to a correct description of the quark mixing angles and masses.  Most of these works did not use the grand unified framework.
Here, we consider a model based on the $SO(10)\otimes Z_2^{\mu-\tau}\otimes Z_2^P$. The first $Z_2$ corresponds to the generalized \mt symmetry. The second $Z_2$ symmetry called \cite{mohparity}  ``parity''  interchanges two  components of the $16$ field transforming as $(4,2,1)$ and $(\overline{4},1,2)$ under the Pati-Salam group decomposition of $SO(10)$. 

Our basic formalism is similar to \cite{mohparity,grimus1,grimus2,yangwang}. 16 dimensional fermions obtain their masses from coupling to three Higgs multiplets transforming as $10,\overline{126}$ and $120$ representations under $SO(10)$. The $SO(10)$ breaking can be achieved with a $210$-plet.  An additional $126$-plet of Higgs is needed in the supersymmetric context to preserve the supersymmetry at the GUT breaking scale. These Higgs  multiplets contain altogether six doublets with quantum numbers of the minimal supersymmetric standard model (MSSM) field $H_d$ and six with that of  $H_u$.
It is assumed that only two appropriate linear combinations of these  Higgs doublets remain light and play the role of the $H_d$ and $H_u$ fields.
This is achieved by the fine tuning conditions \cite{vissani2}.  After this fine tuning, the resulting fermion masses can be written as
\beqa \label{mass term}
-\mathcal{L}_{mass}= \overline{f}_{L} M_{f} f_{R} + \overline{\nu}_{L} M_{D} \nu_{R} + \frac{1}{2} \overline{\nu}_{L} M_{L} \nu_{L}^c + \frac{1}{2} \overline{\nu^{c}_{R}} M_{R} \nu_{R} + h.c. ~,
\eeqa
where $f=u,d,l$ denote the up and down quarks and the charged leptons respectively.  The mass matrices appearing in the above equation can be suitably written  (see \cite{grimus1,grimus2} for details) as
\beqa \label{matrices}
M_d&=& H+F+i~ G~,\nonumber \\
M_u&=&r H+s F+i~ t~ G~, \nonumber\\
M_l&=& H-3 F+i ~p~ G~,\nonumber\\
M_D&=&r H-3s F+i ~q~ G~,\nonumber\\
M_L&=& r_L F~,\nonumber\\
M_R&=& r_R^{-1} F.\eeqa
Here $M_D$ denotes the neutrino Dirac mass matrix. $M_L(M_R)$ is the Majorana mass matrix for the left(right) handed neutrinos which receives contribution only from the vacuum expectation value (vev) of the $\overline{126}$ field. Gauge coupling unification in the minimal model requires that the vev contributing to $M_R$  be close to the GUT scale. The dimensionless parameters $r,s,t,p,q,r_L$ and $r_R$ are determined by the CG coefficients, ratios of vevs and  mixing among the Higgs fields \cite{grimus1}. 

The matrices $H,F,G$ originate from the fermion couplings to the $10,\overline{126}$ and $120$ fields respectively. ($G$) $H,F$  are complex (anti) symmetric matrices in general. However, generalized parity makes them real. In addition, if
all vevs and (hence $r,s,t,p,q,r_L,r_R$) are real then all the Dirac masses in eq.(\ref{matrices}) are Hermitian and $M_L,M_R$ are real.

We assume that the Higgs field in the $10$ and $\overline{126}$ representations are invariant under the generalized \mt symmetry while the $120$ dimensional representation changes sign. This assumption allows spontaneous breaking of the \mt symmetry. The resulting structures for $H,F,G$ are given by
\be \label{hfg} \ba{ccc}
H=\left( \ba{ccc} h_{11}&h_{12}&h_{12}\\h_{12}&h_{22}&h_{23}\\
h_{12}&h_{23}&h_{22}\\ \ea \right);
&
F=\left( \ba{ccc} f_{11}&f_{12}&f_{12}\\f_{12}&f_{22}&f_{23}\\f_{12}&f_{23}&f_{22}\\ \ea \right);
&
G=\left( \ba{ccc}0&g_{12}&-g_{12}\\-g_{12}&0&g_{23}\\g_{12}&-g_{23}&0\\ \ea \right) \ea \ee
All the coefficients in these matrices  are real. They satisfy
\be
\label{yukawasymmetry}
S^T(H,F,G)S=(H,F,-G)~, \ee
where
\be \label{s} S=\left( \ba{ccc} 1&0&0\\
                                                0&0&1\\
                                                0&1&0 \\ \ea \right) \ee
exchanges the second and the third generations.
The effective neutrino mass matrix ${\cal M}_\nu$ for the three light neutrinos  follows from eq.(\ref{mass term}) and eq.(\ref{matrices}):
\be \label{mnu}
{\cal M}_\nu=r_L F-r_R M_D F^{-1} M_D^T \equiv {\cal M}_\nu^{II}+{\cal M}_{\nu}^{I}~. \ee
Here $r_{L,R}$ are inversely related to the vev of the RH triplet component in $\overline{126}$.  This vev may be identified with the GUT scale in the absence of any intermediate scale. In addition, they depend upon the details of the superpotential. Specific expressions for $r_{L,R}$ in the minimal case can be found in  \cite{vissani2, bertolini}. The first term corresponds to the type-II seesaw while the second is the conventional type-I seesaw. In general, both contributions are present but one may dominate over the other. We shall be considering two separate cases corresponding to the type-II and type-I dominance respectively.

The relations $\theta_{23}^{l}=\frac{\pi}{4}$ and $\theta_{13}^{l}=0$ are major predictions and motivation for imposing the \mt symmetry. These can arise if the effective neutrino mass matrix ${\cal M}_{\nu f}$ in flavour basis possesses a \mt symmetry. Let us see how this can come about in our approach. It is easy to see that the fermionic mass matrices in our model satisfy:
\be \label{sq}
S^{-1} M_fS=M_f^* ~, \ee
\be \label{st2} S^{-1}{\cal M}_\nu^{II} S={\cal M}_\nu^{II} ~,\ee
\be \label{st1}S^{-1}{\cal M}_\nu^I S={\cal M}_\nu^{I*}~.\ee 
$f=u,d,l,D$ label the (Dirac) fermionic mass matrices. The ${\cal M}_\nu^{I,II}$ correspond to the type-I and II contributions to the light neutrino mass matrix, eq.(\ref{mnu}). 
Let us note that
\begin{itemize}
\item Eq.(\ref{st2}) implies an exact \mt symmetry for ${\cal M}_\nu^{II}$.
\item  Eqs.(\ref{sq},\ref{st1}) correspond to an invariance under the generalized CP transformation defined \cite{hs,maximal} as 
\be f_{\alpha}\rightarrow i S_{\alpha \beta} \gamma^{0} C \overline{f_{\beta}}^{T} \ee
\item If eq.(\ref{st2}) represents the neutrino masses in flavour basis then
one obtains the predictions $\theta_{23}^{l}=\frac{\pi}{4}$ and $\theta_{13}^{l}=0$. 
\item If eq.(\ref{st1}) holds in the flavour basis then only the $\theta_{23}^{l}$ is maximal with definite correlations of $\theta_{13}^{l}$ with the CP violating phase $\delta_{PMNS}$ \cite{maximal}.
\item  $M_l$ is not diagonal here and hence these predictions do not follow immediately. It is still possible to recover these predictions even with a non-diagonal $M_l$. 
\end{itemize}
Define 
\be \label{ul}  U_l^{\dagger} M_l U_l=D_l  ~,\ee
where $D_l$ is the  diagonal mass matrix for the charged leptons. By factoring out a diagonal phase matrix
$P_{l}$, the $U_l$ can be written as:
\be \label{ul1} U_l\equiv \tilde{U}_l P_l ~\ee 
The neutrino mass matrix in the flavour basis is then given by
\be \label{mnuf} {\cal M}_{\nu f}=P_l^\dagger\tilde{U}_l^\dagger{\cal M}_\nu \tilde{U}_l^\ast P_l^\ast\equiv P_l^\dagger{\cal \tilde{M}}_{\nu f} P_l^\ast ~ \ee
The predictions of the \mt symmetry are recovered if ${\cal \tilde{M}}_{\nu f} $ is \mt invariant. This does not require a diagonal $M_l$. A general  \mt symmetric $\tilde{U}_l$ satisfying $S^{-1}\tilde{U_l}S=\tilde{U}_l$ will do the job in case of the type-II dominance. 
In case of the type-I dominance, one
obtains
$$S^{-1}{\cal \tilde{M}}_{\nu f}S={\cal \tilde{M}}_{\nu f}^*$$
provided $\tilde{U}_l$ also satisfies the same equation. This makes it possible to recover the predictions of the \mt symmetry for a non-diagonal $M_l$ and obtain reasonably good fits to other fermion masses and mixing. 

It is known \cite{hs,maximal}  that  with appropriate choice of  $P_{l}$, $\tilde{U}_l$ can be cast into the following form  if $M_l$ satisfies eq.(\ref{sq}):
\be \label{utildel} \tilde{U}_l=\left( \ba{ccc}
u_{1l}&u_{2l}&u_{3l}\\
w_{1l}&w_{2l}&w_{3l}\\
w_{1l}^*&w_{2l}^*&w_{3l}^* \\ \ea \right) ~,\ee
with real $u_{il}$.
A unitary matrix with this form  can be  parametrized in terms of two angles and a phase. 
\be \label{utildelpara} \tilde{U}_l=P_\eta\left( \ba{ccc}
c_1&s_1c_2&s_1 s_2\\
\frac{s_1}{\sqrt{2}} &-\frac{1}{\sqrt{2}}(c_1 c_2-i\epsilon s_2)&-\frac{1}{\sqrt{2}}(c_1 s_2+i \epsilon c_2)\\
\frac{s_1}{\sqrt{2}} &-\frac{1}{\sqrt{2}}(c_1 c_2+i\epsilon s_2)&-\frac{1}{\sqrt{2}}(c_1 s_2-i \epsilon c_2)\\ \ea \right) ~,\ee
where $\epsilon=\pm 1,s_{1,2}\equiv \sin\theta_{1,2}~,c_{1,2}=\cos\theta_{1,2}$. $c_2$ and $s_2$ can be chosen positive with appropriate choice of $P_l$ in eq.(\ref{ul1}).
$$ P_\eta={\rm diag.} (1,e^{-i\eta},e^{i\eta})$$ is a diagonal phase matrix. The above $\tilde{U_l}$ becomes \mt symmetric if $s_2=c_2$ and $\eta=0$. This defines a one parameter family of the leptonic mass matrices which lead to the prediction
of the \mt symmetry in case of the type-II dominance. We will use this form subsequently in our numerical analysis.

There is an important but unwelcome feature associated with the generalized CP invariance of the mass matrices in eq.(\ref{sq}).
The CKM matrix in this case turns out to be real. To see this explicitly, we note that just as in case of $U_l$, the matrices $U_{u,d}$ diagonalizing the up and down quark masses can be written as $\tilde{U}_{u,d}P_{u,d}$. $\tilde{U}_{u,d}$ have the same form as the RHS of eq.(\ref{utildel}) with the replacement of
$u_{il}$ with $u_{iu,id}$ and $w_{il}$ with $w_{iu,id}$. The phase matrices $P_{u,d}$ can be absorbed in redefining the quark fields and the
remaining part of the CKM matrix  is given by
$$V_{ij}\equiv (\tilde{U}_{u}^\dagger \tilde{U}_{d})_{ij}=u_{iu}u_{jd}+2 Re(w_{iu}w_{jd}^*)$$
which is real since $u_{iu,id}$ are real.

One can generate CP violation in the model by breaking the generalized CP invariance of the mass matrices. This can be done in two ways.  Either one allows complex vev for some of the Higgs doublets as in \cite{grimus1}  or one retains the real vev but allows breaking of the \mt symmetry in the Yukawa couplings. In the following, we will discuss the second alternative.

\section{Fitting fermion spectrum with and without the \mt symmetry}

We now discuss the numerical implications of our model in detail. We assume that either the type-I or the type-II term in the neutrino mass matrix dominates  and carry out analysis separately in each of these two cases. Our input parameters are
$r,s,t,p,q$, eq.(\ref{matrices}) , the real elements of the matrices $G,H,F$, eq.(\ref{hfg}) 
and the overall scales $r_{R,L}$, eq.(\ref{mnu}). Parameter $q$ is absent in the type-II case. An overall rotation $R$ on $G,H,F$: $(G,H,F)\rightarrow R^T(G,H,F)R$  amounts to a choice of initial basis for the 16-plet of fermions. We can use this freedom to set say, $h_{12}=0$. This is done with a specific choice $R=R_{23}^T(\frac{\pi}{4})R_{12}(\theta_{12}^h)R_{23}(\frac{\pi}{4})$. Here $R_{ij}(\theta)$ denotes rotation in the $ij^{th}$ plane by an angle $\theta$ and
$$\tan2 \theta_{12}^h=\frac{2 \sqrt{2} h_{12}}{h_{11}-h_{22}-h_{23}}~.$$ This rotation amounts to redefinition of elements
of $F$ and $G$ which still retain the same form as in eq.(\ref{hfg}).  We continue to use the same notation  for the parameters of the redefined $F,G$. With the choice $h_{12}=0$, we have 14 (15) input parameters in case of type-II (type-I) seesaw dominance. These input parameters together generate 12 fermion masses and six mixing angles. As already remarked, the exact \mt symmetric $H,F,G$ are not able to generate CP violation. We introduce this CP violation by adding a small \mt breaking difference between the 22 and 33 elements in $H$. This one additional parameter now leads to four CP violating phases, one in  the CKM matrix and three in the PMNS matrix.

Our choice of the values of the physical observables  is based on numbers given in \cite{bertolini,grimus2}. We reproduce them here in Table~({\ref{tab:input})  for convenience.
\begin{small}
\begin{table} [h]
\begin{math}
\begin{array}{|c|c||c|c|}
\hline
 m_d &  1.03\pm 0.41&\Delta m^2_{sol}&(7.9\pm 0.3)\times10^{-5} \\ 
m_s&19.6\pm 5.2&\Delta m^2_{atm}&\left(2.2_{-0.27}^{+0.37}\right)\times 10^{-3}\\
m_b & 1063.6_{-86.5}^{+141.4} & \sin  \theta _{12}^q & 0.2243\pm 0.0016 \\
 m_u & 0.45\pm 0.15 & \sin  \theta _{23}^q & 0.0351\pm 0.0013 \\
 m_c&210.3273_{-21.2264}^{+19.0036} & \sin  \theta _{13}^q & 0.0032\pm 0.0005 \\
 m_t&82433.3_{-14768.6}^{+30267.6}& \sin ^2 \theta _{12}^l & 0.31\pm 0.025 \\
 m_e &0.3585\pm 0.0003 &\sin ^2 \theta _{23}^l&0.5\pm 0.065 \\
 m_{\mu }&75.6715_{-0.0501}^{+0.0578} & \sin ^2\theta _{13}^l & < 0.0155 \\
m_{\tau }&1292.2_{-1.2}^{+1.3}&\delta_{CKM}&60^{\circ }\pm 14^{\circ }\\
\hline
\end{array}
\end{math}
\vspace{0.5cm}
\caption{Input values for quark and leptonic masses and mixing angles at $M_{GUT}=2\times 10^{16}$ GeV and $\tan\beta=10$ which we use in our numerical analysis.}
\label{tab:input}
\end{table}
\end{small}

The given numbers for quark masses and mixing correspond to the respective values at the GUT scale obtained from low energy values using MSSM and $\tan \beta=10$. The neutrino masses and mixing that we use are  the low scale values but the effects of 
the evolution to $M_{GUT}$ on the ratio of the solar to atmospheric mass scale and on the mixing angles are known to be small for the normal hierarchical spectrum that we obtain here .  While fitting, we omit the parameters $r_R,r_L$ which define the overall scales of neutrino masses in case of the type-I and type-II seesaw respectively. The ratio of the solar and atmospheric mass scales and neutrino mixing parameters are independent of these overall scales and are used in our definition of $\chi^2$ function instead of the individual neutrino masses. In addition, we assume $\Delta m^2_{atm}$ to be positive corresponding to the normal neutrino mass hierarchy. Parameters $r_R,r_L$ are fixed subsequent to minimization using the atmospheric scale.

\subsection{Numerical analysis: type-II seesaw}
We perform the minimization  in three physically different cases.\\

(A) In this case, we impose the conditions $\theta_{23}^{l}=\frac{\pi}{4}$ and $\theta_{13}^{l}=0$ using a \mt symmetric $\tilde{U}_l$. As discussed in the earlier section,  this is done using parametrization in eq.(\ref{utildelpara}) with $s_2=c_2=\frac{1}{\sqrt{2}}$. The charged lepton mass matrix is then determined completely in terms of three masses and the angle  $\theta_1$.  Using the third of eq.(\ref{matrices}), the real and imaginary parts of $M_l$ can be used to determine respectively elements of   $H$ in terms of that of $F$ and elements of $G$ in terms of p, the charged lepton masses and  $\theta_1$. $f_{12}$ also gets determined in terms of these parameters because of the choice $h_{12}=0$. Thus $f_{22},f_{23},f_{11}, r,s,t,p,\theta_{1}$ are the only free parameters which determine the 11 remaining observables- six quark masses, three angles of the CKM matrix, the solar angle and the solar to atmospheric mass ratio. The $\chi^2$ we minimize is defined in terms of these observables using the values and errors given in Table~(\ref{tab:input}). The result of the minimization are shown in Tables(\ref{tab:2a},\ref{tab:2b}). One obtains a reasonably good fit to all observables except the down and bottom
quark  masses which are respectively $\sim 1.5$ and $\sim 2.5$ sigma away from their  respective mean values.  All other observables are reproduced correctly with very small pulls as seen in the table.\\

(B) In this case, we do not impose the maximality of $\theta_{23}^{l}$ but include $\sin^2\theta_{23}^{l}$  in the  $\chi^2$ to be minimized. $\sin^{2}\theta_{13}^{l}$ is not included in the definition of $\chi^2$ but we require it to be  $\leq 0.0155$ during the minimization. $r,s,t,p$ and elements of $H,F,G$  are now treated as free and the $\chi^2$ definition now includes the charged lepton masses as well. This results in significant improvement in the fit and one is able to fit 15 observables in terms of 13 parameters with $\chi^2=3.01$. The fit to the bottom and the down quark masses also improves. $\delta_{CKM}$ remains zero in this case.\\

(C) For this case, we depart from the exact $23$ symmetry and take $h_{22}$ different from $h_{33}$.  As already discussed, this breaks the generalized CP and results in a non-trivial CKM phase. Remarkably, a very small ($\sim 8\%$) breaking of the $23$ symmetry is able to generate a non-trivial CKM phase and $\chi^2_{min}=3.02$ with 2 degrees of freedom. Bottom quark mass is the only variable which deviates from its central value considerably. \\

\begin{table} [ht]
\begin{small}
\begin{math}
\begin{array}{|c|c|c|c|}
\hline 
&A   &B   &C  \\
\text{\textbf{Quantity}}&\text{\textbf{Pull}} & \text{\textbf{Pull}} &  \text{\textbf{Pull}} \\
\hline
 m_d & -1.47532 	& 0.167255 	& 0.0620115 \\
 m_s & -0.8225		& 0.271662 	& -0.0545523 \\
 m_b & -2.52388		& 1.68787 	& 1.72811 \\
 m_u & 0.274609 	&-0.00446626  	& -0.00184452 \\
 m_c & -0.0125887 	& 0.000159604 	& 0.00744292 \\
 m_t & 0.00190476 	& 0.00901941 	& -0.0199522 \\
 m_e & 0 	& -0.000951761 	& 0.000179815 \\
 m_{\mu } & 0 	& 0.0176266 	& -0.000749102 \\
 m_{\tau } & 0 & -0.0192274 	& -0.017642 \\
 \frac{\Delta m^2_{sol}}{\Delta m^2_{atm}} & 0.679035 & -0.169337
& -0.0544521 \\
 \sin  \theta _{12}^q & -0.0116059 	& 0.00250491 	& -0.00412383 \\
 \sin  \theta _{23}^q &  0.155231 	& -0.00717926 	& 0.0402861 \\
 \sin  \theta _{13}^q & -0.0705362 	& 0.0000163982 	& 0.0163964 \\
 \sin ^2 \theta _{12}^l & 0.112082 	& -0.111783 	& -0.00578002 \\
 \sin ^2 \theta _{23}^l & 0 & 0.129873 	& -0.141465 \\
 \delta _{\text{CKM}} & - 		& - 		& -0.0364271 \\
\hline
 \chi ^2 & 9.80473 & 3.00957 & 3.02019\\
\hline
\hline
\text{\textbf{ }}&\text{\textbf{Predictions}} & \text{\textbf{Predictions}} &  \text{\textbf{Predictions}} \\
\hline
\sin ^2 \theta _{23}^l & 0.5 &- & - \\
\sin ^2 \theta _{13}^l & 0 &  0.000471537 &  0.000226908 \\
\delta _{\text{CKM}} & 0^{\circ} &  0^{\circ} & - \\
\delta _{\text{PMNS}} & 0^{\circ}  & 0^{\circ} & -12.759^{\circ}  \\
\alpha _1 & 180^{\circ} &  180^{\circ} & 169.80^{\circ}  \\
\alpha _2 & 0^{\circ} &  0^{\circ} & -9.445^{\circ}  \\
r_L & 2.8714\times10^{-10}  & 1.8183\times10^{-9} & 1.8645\times10^{-9}\\
\hline
\end{array}
\end{math}
\end{small}
\vspace{0.5cm}
\caption{Best fit solutions for fermion masses and  mixing obtained assuming the type-II seesaw dominance. Various observables
and their pulls obtained at the minimum are shown in three cases (A)-(C) defined in the text.}
\label{tab:2a}
\end{table}

Some of the observables are not part of the $\chi^2$ and their values get fixed at the minimum. These are shown as predictions in Table~(\ref{tab:2a}). These include the CP violating Dirac phase $\delta_{PMNS}$ and the Majorana phases $\alpha_{1,2}$ as defined in \cite{fnt}. These are trivial for the cases (A) and (B) due to the generalized CP invariance but one obtains non-zero values displayed in the Table in case (C).\\

\begin{table}[h]
\begin{small}
\begin{math}
\begin{array}{|c|c|c|c|}
\hline
 \text{\textbf{Parameters}} & A & B & C \\
\hline
 h_{11} & 1.95914 & -0.357916 & -0.818923 \\
 h_{22} & 466.637 & -649.2 &-701.354 \\
 h_{23} & 283.929 & -54.7552 & -32.0485 \\
 h_{33} & 466.637 & -649.2 & -598.783 \\
 f_{11} & -1.25174 & -0.176133 & -0.343138 \\
 f_{12} & 14.2058 & -2.16375 & -2.07269 \\
 f_{22} & -71.54 & 11.5434 & 11.2606 \\
 f_{23} & 95.5358 & -14.754 & -14.3836 \\
 g_{12} & -1.66646 & 3.54811 & 4.19817 \\
 g_{23} & -26.5205 & 614.356 & 617.845 \\
 r & 106.129 & 61.8507 & 61.1056 \\
 s & 114.802 & -109.87 & -121.664 \\
 t & -1.9006 & 67.0199 & 65.9824 \\
 p & 22.8456 & -0.989943 & -0.980791 \\
\hline
\end{array}
\end{math}\end{small}
\vspace{0.5cm}
\caption{Values of parameters of the fermionic mass matrices in eq.(\ref{matrices}) corresponding to the best fit solutions displayed in table(\ref{tab:2a}). The cases(A)-(C) are defined in the text}
\label{tab:2b}
\end{table}
Before going into the more detailed predictions, let us underline some important points connected with the above fits.
\begin{itemize}
\item Detailed fits to fermion masses have been considered in a number of papers with \cite{grimus1, grimus2} or without
\cite{bertolini,splitsusy} the addition of the 120-plet to the minimal $10+\overline{126}$ Higgs fields. The minimal model without the 120-plet  but not imposing reality of the coupling has more parameters than the present case but the fit is not better compared to here, {\it e.g.} the fit in pure type-II
case \cite{bertolini}  with 18 parameters and 15 data points gives a minimum $\chi^2$ around 14.5 \\
\item The  best fit solutions
in  cases (B) and (C) give $\theta_{23}^{l}$ close to maximal and $\theta_{13}^{l}$ close to zero as seen from Table~(\ref{tab:2a}). \\
\item We have fixed the overall scale  of neutrino mass $r_L$ in eq.(\ref{mnu} ) by using the atmospheric scale as normalization. The resulting values are
displayed in Table~(\ref{tab:2a}). In all three cases, $r_L$ comes close to $10^{-10}$.  $r_L$ is related to the mass of the left handed triplet residing in the $\overline{126}$ representation  and to other parameters in the superpotential. Detailed analysis \cite{detailed3,vissani2,bertolini,splitsusy}  has shown that one needs this triplet mass to be at an intermediate scale $\sim 10^{12}$ GeV if the overall neutrino mass scale is to be correctly reproduced.  The presence of such light  triplet conflicts with the gauge coupling unification. An additional $120$-plet does not qualitatively alter the situation. One possible solution suggested \cite{moh54} in the literature is to add a 54-plet of Higgs and allow $SO(10)$ to break first to $SU(5)$ leaving 
a complete 15-plet of Higgs light at around $10^{12}$ GeV.  Other solution corresponds to having split supersymmetry breaking \cite{splitsusy}. Third possibility is to allow type-I seesaw dominance \cite{charan120,charan2}. We shall look at this in the next subsection in the present context.
\end{itemize}

We now turn to predictions in the neutrino sector. The firm predictions of the scheme can be obtained by checking the variation of $\chi^2$ with the values of various observables. As in \cite{bertolini,grimus2} we pin down a specific value  $p_0$ of an observable $P$ by adding a term $$\chi_p^2=\left(\frac{P-p_0}{0.01 p_0}\right)^2$$ to $\chi^2$
and then minimizing
$$\hat{\chi}^2\equiv \chi^2+\chi_P^2~.$$ If $P$ happens to be one of the observables used in defining $\chi^2$ then its contribution is removed from there.
 Artificially introduced small error  fixes the value $p_0$  for $P$ at the minimum of the  $\hat{\chi}^2$.  We then look at the variation of  
\be \label{chibar}
\bar{\chi}^2_{min}\equiv (\hat{\chi}^2 - \chi_p^2) |_{min}
\ee 
 with $p_0$.  The results are displayed in Figs.(1-3). \\

\begin{figure}[h]
 \centering
 \includegraphics[bb=0 0 317 203]{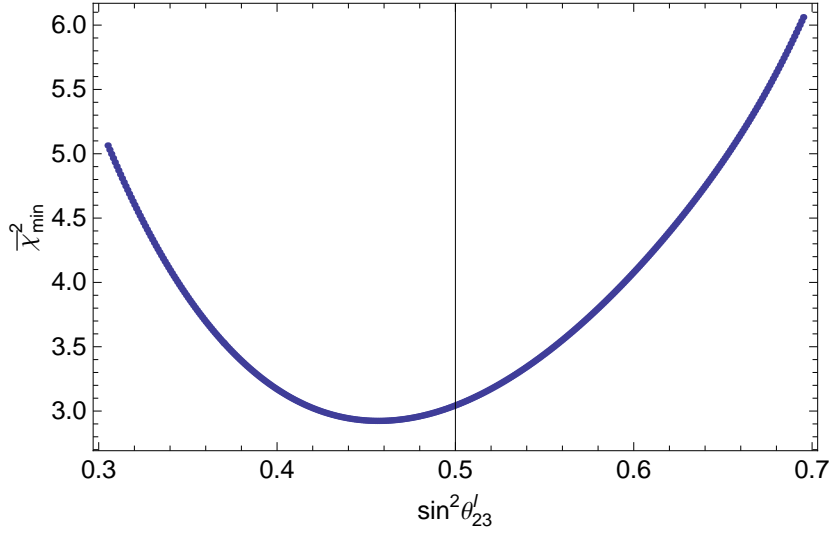}
 \caption{Variation of $\bar{\chi}^2_{min}$ with $\sin^2\theta_{23}^{l}$ in Type-II seesaw.}
 \label{fig:1}
\end{figure}

\begin{figure}[ht]
 \centering
 \includegraphics[bb=0 0 320 208]{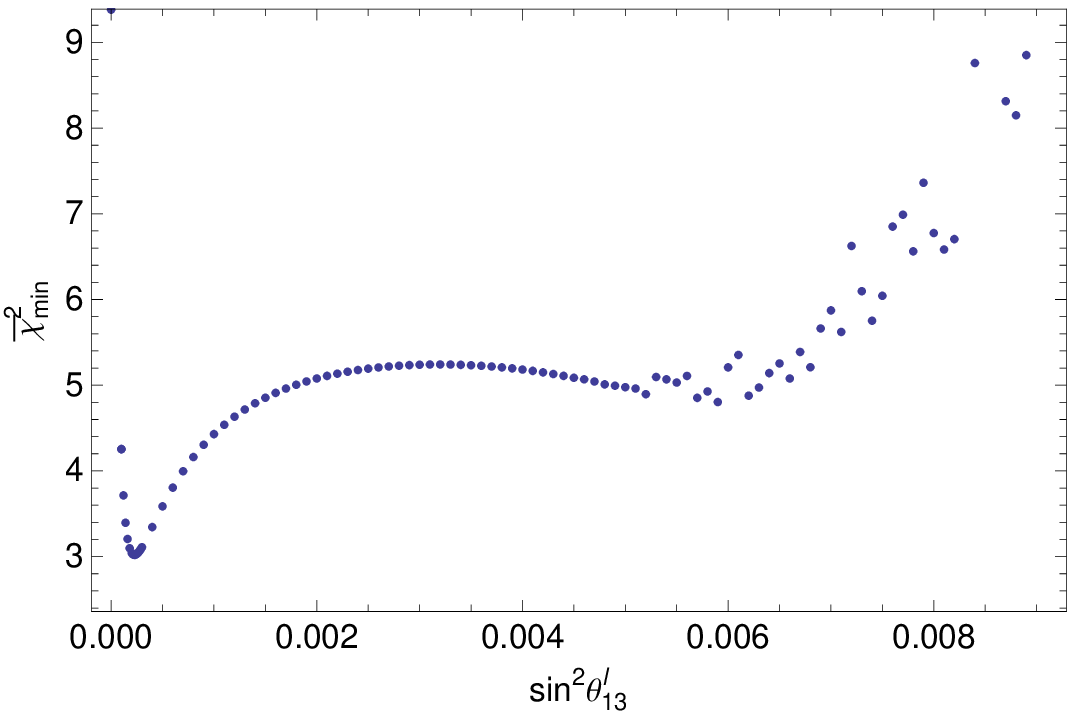}
 \caption{Variation of $\bar{\chi}^2_{min}$ with $\sin^2\theta_{13}^{l}$ in Type-II seesaw.}
 \label{fig:2}
\end{figure}

\begin{figure}[ht]
 \centering
 \includegraphics[bb=0 0 315 208]{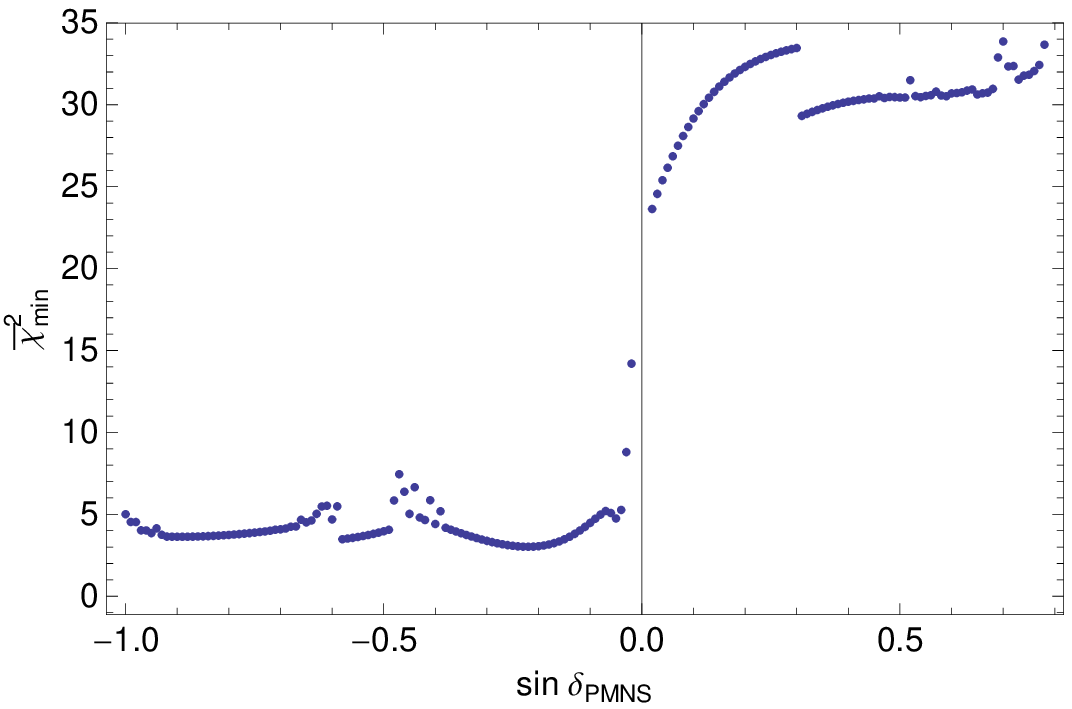}
 \caption{Variation of $\bar{\chi}^2_{min}$ with $\sin\delta_{PMNS}$ in Type-II seesaw.}
 \label{fig:3}
\end{figure}

Fig.(\ref{fig:1}) shows the variation of $\bar{\chi}_{min}^2$ for various pinned downed values of $\sin^2\theta_{23}^{l}$. It is seen that the minimum occurs when $\sin^2\theta_{23}^{l}$ is fixed to around $~0.46$ rather than the value $~0.5$ obtained in the fits shown in Table~({\ref{tab:2a}). The variation of $\bar{\chi}_{min}^2$ is not drastic and all values in the range $0.3-0.7$ are allowed at 90$\% $CL. In comparison, variation of $\bar{\chi}_{min}^2$ with $\sin^2\theta_{13}^{l}$ shown in Fig.(\ref{fig:2}) is little more significant. There is a preference for values close to zero but values up to 0.008 cannot be ruled out at 90$\%$ confidence level. Fig.(\ref{fig:3}) shows the prediction for the PMNS phase in the leptonic mixing matrix.  Clear prediction is the negative values for the $\sin\delta_{PMNS}$. However, all negative values are allowed within the 90$\%$ confidence limit. \\

\subsection{Numerical analysis: type-I seesaw}
The structure of the neutrino mass matrix in the type-I case is qualitatively different compared to the type-II case. Unlike ${\cal M}_\nu^{II}$, ${\cal M}_\nu^I$ is not \mt invariant in general. But it can be made approximately \mt symmetric if either $120$ contribution or the $10+\overline{126}$
dominates in $M_D$, see eq.(\ref{matrices}). We discuss below fits in three qualitatively different cases as done for the type-II dominance.\\ 

(A) Here we impose the exact \mt symmetry for ${\cal M}_\nu^{I}$ by hand, i.e. by choosing $q=0$ in $M_D$. As before, $U_l$ is also chosen \mt symmetric. The input parameters and observables are the same as in the case (A) of type-II seesaw. The results of the fits are displayed in the first column of the Table~(\ref{tab:3a}). The total $\chi^2$ involves 11  observables and is determined by 8 parameters . The minimum value is $\sim 13$.  While most observables can be fitted nicely, the top quark mass  deviate by $3.6\sigma$ from the central value. Enforcing the exact \mt symmetry does not appear to be a very good choice.\\

(B) In this case, we do not take $q=0$. ${\cal M}_\nu^I$ now satisfies  eq.(\ref{st1})   and is  not symmetric under \mt symmetry.
    $\theta_{23}^{l}$ is not fixed to be maximal but is included in the definition of $\chi^2$. As in the earlier case (B), $\chi^2$ is defined by 15 observables and is determined in terms of 14 parameters. The CP violating phases are zero in this case and the CKM phase is therefore not included in $\chi^2$.  Experimental bound on $\theta_{13}^{l}$ shown in Table~(\ref{tab:input}) is imposed during  the minimization. One now gets excellent fit to all the included variables with $\chi^2_{min}=0.017$.\\

(C) In this  case  we introduce a small explicit \mt symmetry breaking by assuming $h_{22}\not =h_{33}$ in eq.(\ref{matrices}). This allows CP violation. $\chi^2$ definition now includes all 16 observables and depends on 15 parameters. Bound on $\theta_{13}^{l}$ is imposed during minimization. Once again we get an excellent fit to all the observables with $\chi^2_{min}=0.18$. CP violating phases in the PMNS matrix come as predictions.\\
\begin{small}
\begin{table}[ht]
\noindent\(
\begin{array}{|c|c|c|c|}
\hline 
 &A   &B   &C  \\
   \text{\textbf{Quantity}} & \text{\textbf{Pull}} & \text{\textbf{Pull}} &  \text{\textbf{Pull}} \\
\hline
 m_d  & -0.31569  & 0.0346007  & -0.379829 \\
 m_s  & 0.473034 & -0.0483779 & -0.0717277 \\
 m_b  & -0.108264 & -0.113763 & -0.114314 \\
 m_u & 0.50263 & 0.00026323 & 0.00344698 \\
 m_c & -0.151225 & -0.000606809 & -0.00938266 \\
 m_t & -3.60744 & -0.0193107 & 0.0122663 \\
 m_e & 0 & -4.874\times10^{-6} & 0.0000348858 \\
 m_{\mu } & 0 & 0.000480511 & 0.00078371 \\
 m_{\tau } & 0 & 0.00254153 & -0.0106065 \\
 \frac{\Delta m^2_{sol}}{\Delta m^2_{atm}}  & -0.00977627 &
-0.00192856 & 0.0125218 \\
 \sin  \theta _{12}^q &0.0218205 &  -0.00061312 & 0.00761817 \\
 \sin  \theta _{23}^q & 0.00289271 & 0.00129946 & 0.0284214 \\
 \sin  \theta _{13}^q & -0.238953 & -0.00823361 & 0.0366413 \\
 \sin ^2 \theta _{12}^l & -0.0129712 & 0.000590904 & -0.00265193 \\
 \sin ^2 \theta _{23}^l & 0 & -0.00544523 & 0.0289959 \\
 \delta _{\text{CKM}} & - & - & -0.120278 \\
\hline
 \chi ^2 & 13.6821 & 0.0169632 &0.180526\\
\hline
\hline
\text{\textbf{ }}&\text{\textbf{Predictions}} & \text{\textbf{Predictions}} &  \text{\textbf{Predictions}} \\
\hline
\sin ^2 \theta _{23}^l & 0.5 &- & - \\
 \sin ^2 \theta _{13}^l & 0  & 0.0135605 & 0.013505\\
 \delta _{\text{CKM}} & 0^{\circ} & 0^{\circ}  & -\\
 \delta _{\text{PMNS}} & 0^{\circ} & 0^{\circ} & -0.287748^{\circ} \\
 \alpha _1 & 180^{\circ} & 0^{\circ} & 2.156^{\circ}  \\
\alpha _2 & 0^{\circ} &  0^{\circ} & 2.616^{\circ}  \\
r_R & 4.1143\times10^{-11}  & 5.2329\times10^{-18} & 5.0093\times10^{-18}\\
\hline
\end{array}
\)
\vspace{0.5cm}
\caption{Best fit solutions for fermion masses and  mixing obtained assuming the type-I seesaw dominance. Various observables
and their pull obtained at the minimum are shown in three cases (A)-(C) defined in the text.}
\label{tab:3a}
\end{table}
\end{small}

Noteworthy features of the fits in (B) and (C) cases above are the following:
\begin{itemize}
\item The overall neutrino mass scale is determined to be around $r_R\sim 5\times 10^{-18}$.  $r_R$ is related to the ratio of the vev of the doublet and the RH triplet components in $\overline{126}$. The values of $r_R$ obtained here are similar to the values obtained in \cite{grimus2} which assume $\overline{126}$ RH triplet vev to be at the GUT scale. Thus one does not need an intermediate scale in order to fit the neutrino masses and one can obtain the gauge  coupling unification. This is consistent with observations in \cite{charan2,grimus1, grimus2}.\\
\item Maximality of $\theta_{23}^{l}$ is not imposed. But it is fixed to be very close to $\frac{\pi}{4}$ at the minimum in both the cases. The departure from the \mt symmetry results in $\theta_{13}^{l}$ being non-zero and is fixed around the upper bound at the minimum as seen from the Table~(\ref{tab:2a}).\\
\item Although an explicit breaking of the \mt symmetry is introduced in case (C), the amount of the breaking required in order to obtain the large CP violating phase is extremely  tiny,
\be
\label{23breaking}
 \frac{h_{22}-h_{33}}{h_{22}+h_{33}}\sim 0.0045 ~.\ee
\item The exact \mt symmetry is known \cite{23as} to lead to the unwanted predictions $V_{ub}=V_{cb}=\sin^2\theta_{23}^{l}=0$.
Here we have two sources of breaking this symmetry, spontaneous through the vev of the 120-plet and explicit through 
eq.(\ref{23breaking}) which allows one to reproduce the mixing angles correctly. In spite of the \mt breaking, the final fermion
mass matrices display a remarkably good \mt symmetry. We make this explicit by giving the quark and lepton mass matrices 
in the case (C) above in Appendix (A). $M_{u,d,l}$ and ${\cal M}_\nu^I$ are seen to be nearly \mt symmetric. There is an order of magnitude difference in the imaginary parts of the 12 and 13 elements of ${\cal M}_\nu^{I}$.  But these imaginary parts are much smaller than the corresponding \mt symmetric real parts. The only source of the large \mt breaking occurs as a difference between the 12 and 13 elements of the  Dirac neutrino mass matrix $M_D$. This  results from the spontaneous breakdown and rather large value of the parameter $q$.
\item As in \cite{grimus1,grimus2} we have concentrated here in obtaining generic fits to fermion masses rather than considering the entire parameter space of the theory given by the Yukawa couplings and basic parameters in the superpotential.  Parameters in fermion mass matrices are related to the strengths of the light Higgs components in various $SO(10)$ Higgs representations. These are determined by the fine tuning conditions and the full superpotential. Grimus and K\"{u}hb\"{o}ck \cite{grimus1} have laid down consistency constraints on these parameters following from  these
fine tuning relations  and from the requirement that the Yukawa couplings stay in the perturbative regime. We have checked that these conditions are satisfied by the parameters given in the Table~(\ref{tab:2b},\ref{tab:3b}).
\end{itemize}
\begin{small}
\begin{table}[ht]
\begin{math}
\begin{array}{|c|c|c|c|}
\hline
 \text{\textbf{Parameters}} & A & B & C \\
\hline
 h_{11} & 907.294 & 34.9749 & 35.0178 \\
 h_{22} & 119.541 & 554.305 & 556.777 \\
 h_{23} & -119.052 & 554.457 & 554.429 \\
 h_{33} & 119.541 & 554.305 & 551.775 \\
 f_{11} & 74.5214 & -15.7284 & -15.716 \\
 f_{12} & -2.82327 & 20.8852 & 20.8951 \\
 f_{22} & -74.237 & -29.4577 & -29.4636 \\
 f_{23} & 74.2104 & -29.5265 & -29.5305 \\
 g_{12} & 182.676 & 3.10944 & 2.79728 \\
 g_{23} & -4.5309 & -3.5854 & -3.21385 \\
 r & 1.24579 & 83.0642 & 83.7973 \\
 s & 0.266298 & 176.883 & 178.571 \\
 t & 0.844656 & 0.450978 & 1.0715 \\
 p & 2.35413 & 0.0117737 & 0.011244 \\
 q & 0 & 4042.93 & 4537.34\\
\hline
\end{array}\end{math}
\vspace{0.5cm}
\caption{Values of parameters of the fermionic mass matrices in eq.(\ref{matrices}) corresponding to the best fit solutions displayed in table(\ref{tab:3a}). The cases (A)-(C) are defined in the text}
\label{tab:3b}
\end{table}
\end{small}

We follow a similar procedure as in the type-II case to obtain possible predictions on the neutrino mixing variables. We pin down an observable P to a specific value $p_0$ by adding a contribution $\chi_P^2$  to $\chi^2$. We then determine the variation of $\bar{\chi}_{min}^2$ defined earlier with $p_0$. Variations of $\bar{\chi}_{min}^2$ obtained at different local
minima are shown as scattered plots in Fig.(4-6). \\

\begin{figure} [h]
 \centering
 \includegraphics[width=7cm,height=11cm,angle=270,bb=0 0 504 720]{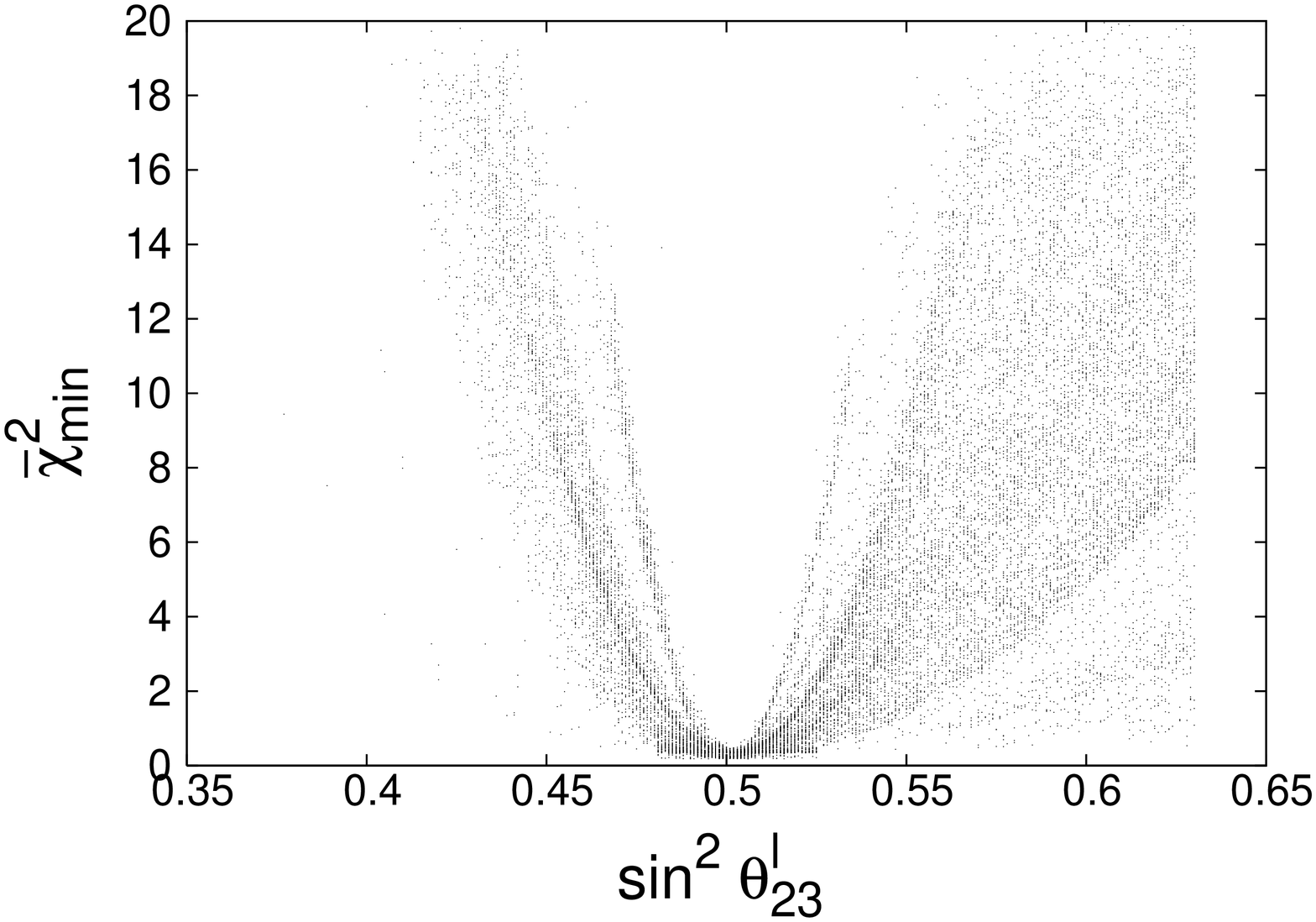}
\vspace{1cm} 
\caption{Variation of $\bar{\chi}^2_{min}$ with $\sin^2\theta_{23}^{l}$ in Type-I seesaw.}
 \label{fig:4}
\end{figure}

\begin{figure}[h]
 \centering
 \includegraphics[width=7cm,height=11cm,angle=270,bb=0 0 504 720]{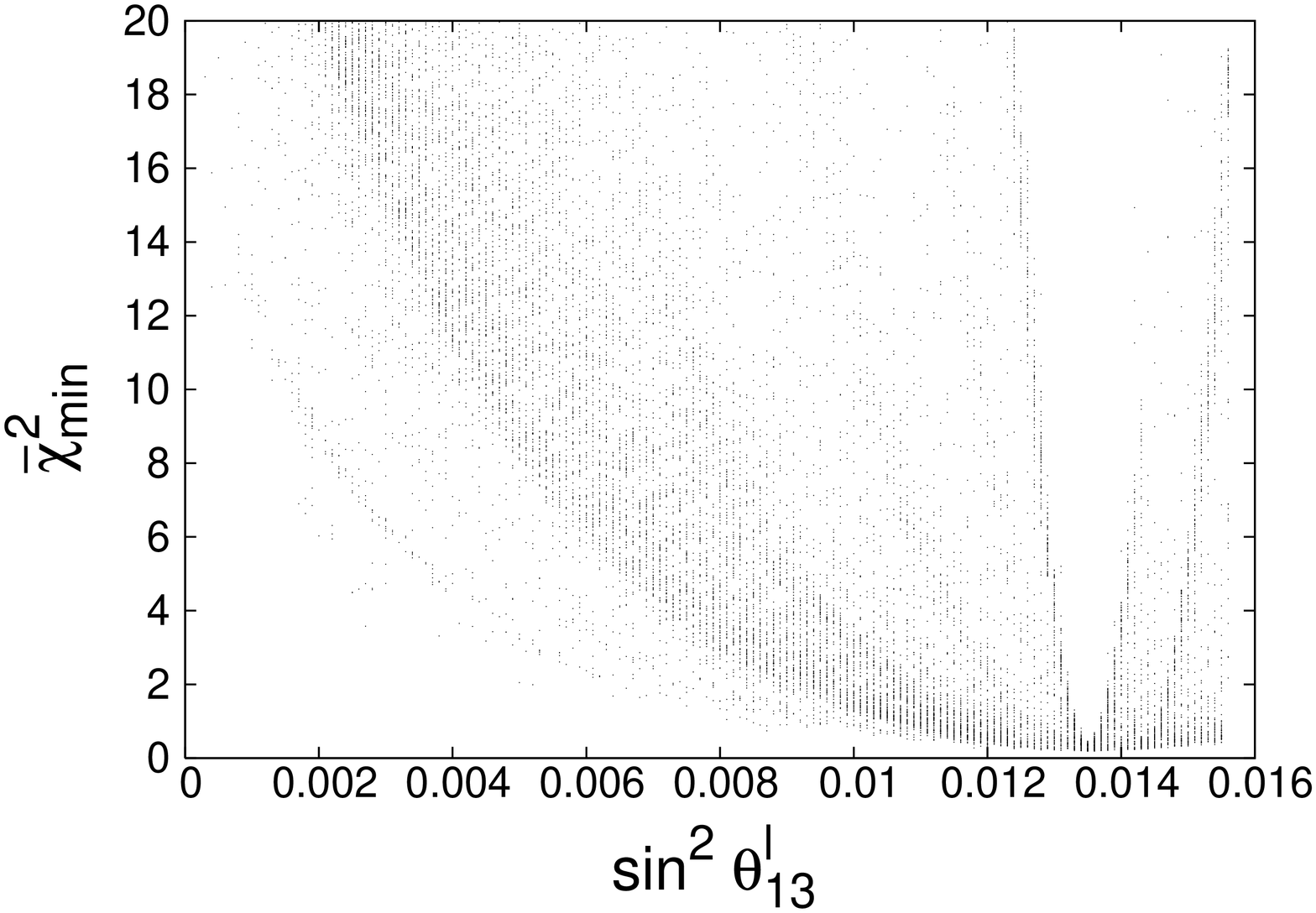}
\vspace{1cm} 
 \caption{Variation of $\bar{\chi}^2_{min}$ with $\sin^2\theta_{13}^{l}$ in Type-I seesaw.}
 \label{fig:5}
\end{figure}

\begin{figure}[h]
 \centering
 \includegraphics[width=7cm,height=11cm,angle=270,bb=0 0 504 720]{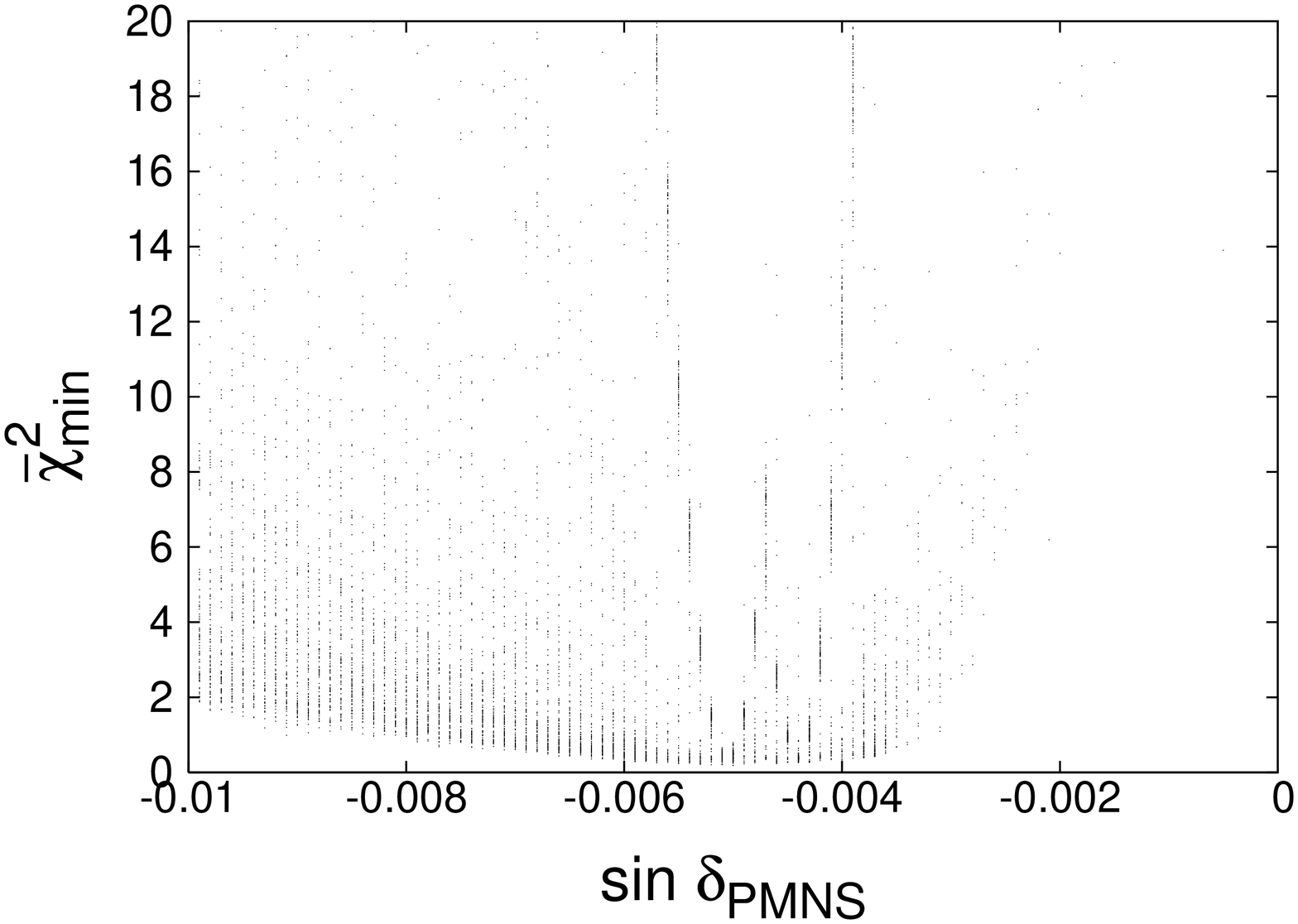}
\vspace{1cm} 
 \caption{Variation of $\bar{\chi}^2_{min}$ with $\sin\delta_{PMNS}$ in Type-I seesaw.}
 \label{fig:6}
\end{figure}

Clear predictions emerge unlike in the type-II case. As Fig.(\ref{fig:4}) shows, the $\sin^2\theta_{23}^{l}$ is preferentially restricted near $0.5$ and one obtains the limit $\sim 0.42-0.63$ at the 90$\%$ CL. Fig.(\ref{fig:5}) shows similar variation with respect to  $\sin^2\theta_{13}^{l}$. Here, the preferred values occur near the present limit and one obtains $\sin^2\theta_{13}^{l}>0.005$ at 90$\%$ CL. The predicted values for $\sin\delta_{PMNS}$  are displayed in Fig.(\ref{fig:6}). These are negative but very small. 

All the above solutions are obtained through an extensive  search using  the random search algorithm in  Mathematica and 
the MINUIT sub-routine in FORTRAN and we have shown in tables the solutions corresponding to the minimum $\chi^2$ that we obtained. Considering the non-linearity and complexity of the problem here, it is difficult to rule out the existence of still lower minima and predictions  may improve if they exist.

We end this section with a comment on the specific \mt symmetry defined by $S$ used in eq.(\ref{s}). Definition of $S$ is basis-dependent.  One could change the original basis of the 16-plet through an arbitrary rotation $R$.  The structure of the Yukawa couplings and the resulting  fermionic mass matrices would look different in the new basis. The new Yukawa couplings would  still  satisfy the same equation as (\ref{yukawasymmetry}) but now with a rotated $S$: $S_R\equiv R^TSR$. Thus the \mt symmetry may appear to look different with different choices of $R$. Specifically, if $R$ corresponds to a rotation by $\frac{\pi}{4}$ in the 23 plane then the $S_R$ assumes the form
\be \label{sr} S_R=\left( \ba{ccc} 1&0&0\\
                                                0&-1&0\\
                                                0&0&1 \\ \ea \right) ~.\ee
This is nothing but the $Z_2$ symmetry imposed in \cite{grimus1} which is  thus equivalent to  the  generalized \mt symmetry considered here if both remain unbroken. Difference arises after these symmetries are broken. 
Ref. \cite{grimus1} uses complex vev to achieve $Z_2$ breaking as a result of which analogue of eqs.(\ref{sq}-\ref{st2}) do not hold in their case.
In our approach, we introduce small explicit breaking of   \mt  symmetry in $H$.   The model in \cite{grimus1} has 20 free parameters compared to 15  used here.

Note that the  explicit breaking of the \mt symmetry is technically natural in the supersymmetric context.  Alternatively, one can achieve such breaking by introducing an additional 10-plet of the Higgs field which changes sign under the \mt symmetry. Combined contributions of these two 10-plets would then give an explicitly \mt non-invariant  $H$.
 \section{Summary}  Aim of this paper was to integrate the successful $\mu$-$\tau$ symmetry within the $SO(10)$ framework in order to obtain a constrained  picture of fermion masses and theoretical understanding of the largeness of the atmospheric mixing angle. The explicit model discussed here provides this integration rather well as shown by the  detailed fits to fermion masses presented in  Tables~({\ref {tab:2a},\ref{tab:3a}).
Interestingly, mass matrices obtained in the model under consideration display a generalized CP invariance if Yukawa couplings are taken to be \mt symmetric. Small explicit breaking of this symmetry is sufficient to generate the required CP violating phase.
The best scenario is obtained in the type-I seesaw model with very tiny explicit \mt symmetry breaking. This scenario is characterized by the predictions $\sin^2\theta_{23}^{l}\sim 0.42-0.63$, $\sin^2\theta_{13}^{l}>0.005$ and negligible CP violation in neutrino oscillations.
Final quark, the charged lepton and the light neutrino  mass matrices (collected in the Appendix)  respect  \mt symmetry to a very good approximation indicating that this symmetry provides a good description of the entire fermion spectrum rather than being restricted to the neutrino sector alone.
\section{Acknowledgments}
We thank C. S. Aulakh for useful discussions.
\section{Appendix (A)}
We list here the fermion mass matrices following from eq.(\ref{matrices}) using the best fit values of the parameters given in Table~(\ref{tab:3b}) corresponding to the type-I seesaw mechanism. The neutrino mass matrix is expressed in eV units while all other  mass matrices are expressed in MeV units.
\begin{small}
\be
M_d=\left(
\begin{array}{ccc}
 19.3018 & 20.8951+2.79728 i & 20.8951-2.79728 i \\
 20.8951-2.79728 i & 527.314 & 524.898-3.21385 i \\
 20.8951+2.79728 i & 524.898+3.21385 i & 522.311
\end{array}
\right) \ee 
\be 
M_u= \left(
\begin{array}{ccc}
 127.971 & 3731.25+2.99727 i & 3731.25-2.99727 i \\
 3731.25-2.99727 i & 41395.1 & 41186.3-3.44363 i \\
 3731.25+2.99727 i & 41186.3+3.44363 i & 40975.9
\end{array}
\right)\ee
\be 
M_l=\left(
\begin{array}{ccc}
 82.1659 & -62.6852+0.0314526 i & -62.6852-0.0314526 i \\
 -62.6852-0.0314526 i & 645.168 & 643.02-0.0361365 i \\
 -62.6852+0.0314526 i & 643.02+0.0361365 i & 640.166
\end{array}
\right) \ee
\be
M_D=\left(
\begin{array}{ccc}
 11353.7 & -11193.7+12692.2 i & -11193.7-12692.2 i \\
 -11193.7-12692.2 i & 62440.4 & 62279.4-14582.3 i \\
 -11193.7+12692.2 i & 62279.4+14582.3 i & 62021.3
\end{array}
\right) \ee
\be
{\cal M}_\nu^{I}=\left(
\begin{array}{ccc}
 -0.0242264 & -0.0143681+0.0004742 i & -0.0143657-0.0000755678 i \\
 -0.0143681+0.0004742 i & -0.0128288+0.00678282 i & -0.0163109+0.000214216 i \\
 -0.0143657-0.0000755678 i & -0.0163109+0.000214216 i & -0.0127693-0.00629525 i
\end{array}
\right)
 \ee
\end{small}

\end{document}